# Experimental indications of the 3+1 neutrino model with one sterile neutrino

A. Serebrov, R. Samoilov, M. Chaikovskii

*Petersburg Nuclear Physics Institute, National Research Center Kurchatov Institute, 188300, Orlova roscha 1, Gatchina, Russia*

E-mail: serebrov_ap@pnpi.nrcki.ru

## Abstract

The possibility of the validation of the 3+1 neutrino model is considered in the context of the new result of the Neutrino-4 experiment, the direct observation of the oscillation effect at parameter region $\Delta m_{14}^2 = (7.3 \pm 0.13_{st} \pm 1.16_{syst})$ eV$^2$ and $\sin^2 2\theta_{14} = 0.36 \pm 0.12_{stat}(2.9\sigma)$, also using LSND anomaly, MiniBooNE anomaly, reactor antineutrino anomaly, and gallium anomaly observed in experiments with radioactive sources GALLEX/GNO, SAGE and BEST. We analyze agreement of the neutrino parameter values obtained in the Neutrino-4 experiment with the results of other reactor type experiments NEOS, DANSS, STEREO, PROSPECT, the experiments at accelerators MiniBooNE, LSND, and the IceCube experiment. We present the analysis of the 3+1 neutrino model which demonstrates the agreement among the experiments within current experimental accuracy. Also, current accuracy allows us to make estimations of the effective masses of the neutrinos. The mass of the sterile neutrino obtained in the Neutrino-4 experiment (in assumption $m_4^2 \approx \Delta m_{14}^2$) is $m_4 = (2.70 \pm 0.22)$eV. Using the estimations of the mixing angles we calculated the values of the electron, muon and tau neutrinos: $m_{4\nu_e}^{\text{eff}} = (0.82 \pm 0.16)$eV, $m_{\nu_\mu}^{\text{eff}} = (0.41 \pm 0.25)$eV, $m_{\nu_\tau}^{\text{eff}} \leq 0.60$eV. These results are compared with the values obtained in the experiments in the direct measurement of the neutrino mass KATRIN and GERDA. The Pontecorvo–Maki–Nakagawa–Sakata matrix with four states and the mixing scheme of the flavor neutrinos with the sterile state are presented.

## 1. Introduction

In Standard Model there are only three massless neutrinos which take part in the weak interaction. The measurements of the total decay probability of the Z-boson restrict the number of such particles. The current value is $N\nu = 2.984 \pm 0.008$. However, the experimentally observed neutrino oscillations require neutrinos to have non-zero mass. Moreover, the anomalies are observed in various experiments at reactors and accelerators: LSND anomaly at CL 3.8$\sigma$ [1], MiniBooNE anomaly at CL 4.7$\sigma$ [2], the reactor antineutrino anomaly (RAA) at 3$\sigma$ CL [3,4], and gallium anomaly in experiments GALLEX/GNO and SAGE at the CL 3.2$\sigma$ [5-7]. The straight way to adjust theory is to extend the SM by including the masses of the neutrinos and adding sterile neutrinos to the particle model. One of the possible extensions is the 3+1 neutrino model with one sterile state. Also, the sterile neutrino is considered to be a candidate to the particle of the Dark Matter.

Previously, the comparison of the results of the Neutrino-4 and the results of the listed above experiments was performed in our work [8] as a part of the analysis of the Neutrino-4 experiment. This work is focused on the experimental verification of the 3+1 neutrino model using the latest results of the Neutrino-4 experiment, recently published in the work [9].

## 2. The results of the Neutrino-4 experiment

The Neutrino-4 experiment [9] is carried out at the SM-3 reactor (Dimitrovgrad, Russia). The experiments at the reactors where one compares the measured antineutrino flux from the reactor with the calculated expected value have obvious disadvantage – it relies on the accuracy of the estimation of the expected flux and the detailed information about the efficiency of a neutrino detector. The main advantages of the Neutrino-4 experimental setup are the compact reactor core (35x42x42cm$^3$) of the SM-3 reactor, with high reactor power equal to 90 MW and the application of the method of the relative measurements carried out by the moveable detector. In this experiment we observe the oscillation process in form of the flux as a function of the parameter L/E according to the expression (1).

$$P(\bar{\nu}_e \to \bar{\nu}_e) = 1 - \sin^2 2\theta_{14} \sin^2\left(1.27\frac{\Delta m_{14}^2[\text{eV}^2]L[\text{m}]}{E_{\bar{\nu}}[\text{MeV}]}\right) \quad (1),$$

where $E_{\bar{\nu}}$ is the antineutrino energy in MeV, L is the distance in meters, $\Delta m_{14}^2$ is the difference between squared masses of electron and sterile neutrinos, $\theta_{14}$ is the mixing angle of electron and sterile neutrinos. The experimental test of the oscillation hypothesis requires measurements of the antineutrino flux and spectrum as near as possible to a practically point-like antineutrino source. The detailed description of the Neutrino-4 detector, the preparation of the neutrino laboratory, the measurements of the background, and the scheme of data taking by the full-scale detector can be found in the ref. [9]. Below we demonstrate the method and the results of the direct observation of the oscillations in the Neutrino-4 experiment.

The measurements of the antineutrino flux are carried out by comparing the signals obtained with operating and shut down reactor. The difference between the measured signals is the required antineutrino flux, while the cosmic background, which is the main problem for the experiment, is cancelled out. The measured antineutrino flux as a function of the distance and energy is convenient to present in the form of the matrix with elements - $N(E_i, L_k)$ [9].

The well-known problem of the discrepancy between the calculated spectrum and the measured one was also observed in our experiment [9]. Therefore, the data analysis should be performed in way that minimize the influence of the expected energy spectrum. We utilize the equation (2) to perform the model independent analysis. The numerator in (2) is the measured number of the events multiplied by the geometric factor $L^2$, and the denominator is the expected number of the neutrino events averaged over whole set of distances. The spectral dependence is cancelled out in this ratio.

$$R_{ik}^{\text{exp}} = N(E_i, L_k)L_k^2/K^{-1}\sum_k^K N(E_i, L_k)L_k^2 = \frac{1-\sin^2 2\theta_{14} \sin^2(1.27\Delta m_{14}^2 L_k/E_i)}{K^{-1}\sum_k^K(1-\sin^2 2\theta_{14}\sin^2(1.27\Delta m_{14}^2 L_k/E_i))} = R_{ik}^{\text{th}} \quad (2)$$

In the set of distances which are much greater than the oscillation length the denominator can be simplified:

$$R_{ik}^{\text{th}} \approx \frac{1-\sin^2 2\theta_{14}\sin^2(1.27\Delta m_{14}^2 L_k/E_i)}{1-\frac{1}{2}\sin^2 2\theta_{14}} \xrightarrow[\theta_{14}=0]{} 1 \quad (3)$$

One can see that the equation (3) is the oscillation curve, and it differs from the equation (1) only by the constant $1 - 1/2 \sin^2 2\theta_{14}$ in the denominator. Hence the oscillation process can be extracted from the experimental data. The comparison of the experimental results with the matrix obtained in MC simulation can be carried out using $\Delta\chi^2$ method, i.e. the expression $\sum_{i,k}(R_{ik}^{exp} - R_{ik}^{th})^2/(\Delta R_{ik}^{exp})^2 = \chi^2(\sin^2 2\theta_{14}, \Delta m_{14}^2)$. The results of the data analysis is shown in fig. 1a. The oscillation length for the average energy is 1.4m, i.e. less than the width of the biological shielding of the reactor. Nevertheless, the oscillation process can be observed by the described method. The result of the analysis using $\Delta\chi^2$ method is shown in fig.1b. The area of oscillation parameters colored in pink are excluded with CL more than 3σ. However, in area $\Delta m_{14}^2 = 7.3 eV^2$ and $\sin^2 2\theta_{14} = 0.36 \pm 0.12_{stat}$ the oscillation effect is observed at 2.9 σ CL. Here we also illustrate the areas of Reactor anomaly and Gallium anomaly.

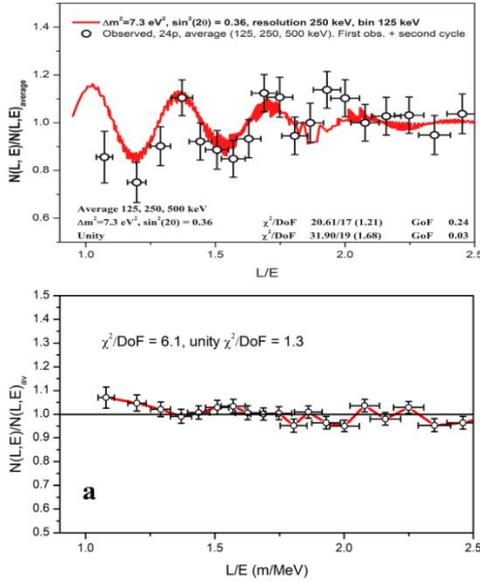
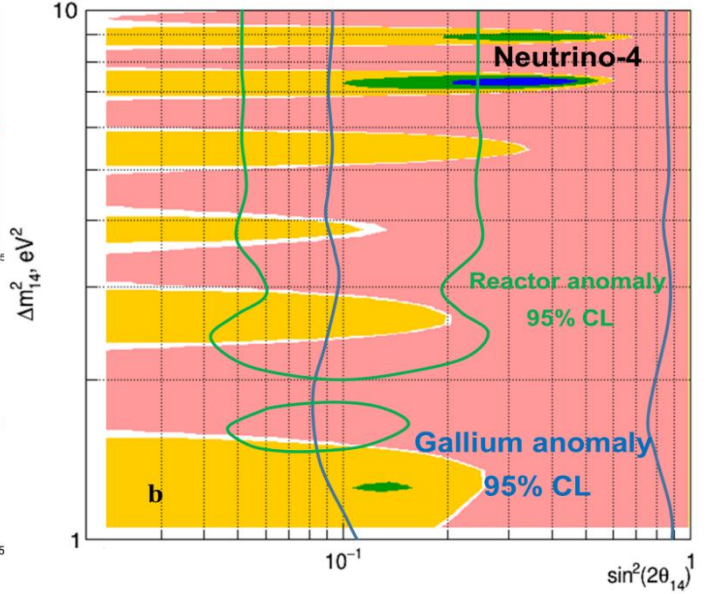

Fig.1 a – The oscillation curve of the neutrino signal reactor ON and the measurements of the background with reactor OFF, b – The results of the analysis of data. on the plane $\Delta m_{14}^2, \sin^2 2\theta_{14}$.

**3. The comparison of the Neutrino-4 results with reactor antineutrino anomaly, gallium anomaly, and the results of the KATRIN and GERDA experiments**

The Neutrino-4 experiment is aimed at the direct measurement of the oscillation parameter $\sin^2 2\theta_{14}$, which has value two times bigger than the neutrino flux deficit at large distance. Therefore, in order to compare the results of the Neutrino-4 experiment with the RAA and GA the value of $\sin^2 2\theta_{14}$ can be used to derive the deficit or vice versa. We prefer to use the value of the oscillation parameter $\sin^2 2\theta_{14}$ in our comparative analysis.

The deficit of the neutrino in gallium experiments considering new result of the BEST experiment is $0.84 \pm 0.04$ [10] and corresponding $\sin^2 2\theta_{14} \approx 0.32 \pm 0.08(4\sigma)$. This result confirms result of the Neutrino-4 experiment with $\sin^2 2\theta_{14} \approx 0.36 \pm 0.12(2.9\sigma)$. The combined value of the Neutrino-4 and new GA value is $\sin^2 2\theta_{14} \approx 0.33 \pm 0.07 (4.9\sigma)$, and it is 2.5 times higher than the RAA deficit, while the Neutrino-4 and GA values are close $0.36 \pm 0.12$ and $0.32 \pm 0.08$ correspondingly. Also, the interpretation of the RAA measurements is based on the calculated spectrum which still has the unexplained tensions with the measured one. It should be noted, that RAA is based on a complex and quite difficult method of absolute measurements, but Neutrino-4 experiment and BEST experiment use method of relative measurements, which is more reliable. Therefore, we use the sterile neutrino parameters $\sin^2 2\theta_{14} \approx 0.33 \pm 0.07$ and $\Delta m_{14}^2 = (7.3 \pm 0.13_{st} \pm 1.16_{syst})$ eV$^2$ in our further analysis.

The discrepancy between the area of the RAA and the area of the BEST result combined with the previous SAGE and GALLEX results is obvious and requires possible interpretation. It should be mentioned that it was noted in [11] that an alternative interpretation of the uncertainty of the spectra is possible. This is not a bump in the 5 MeV area, but a hole in the maximum spectrum in the 3 MeV area ("we should discuss not the" bump "in 5 MeV area, but the" hole "in 3 MeV area."[11]). With this approach, this additional deficit in the region of the spectrum maximum is approximately 10%, as shown in [12] in Fig. 5. As a result, the total deficit can reach 17%, which corresponds to $\sin^2 2\theta_{14} \approx 0.34$. This approach allows the results of the RAA and the BEST experiment to be reconciled. The presented estimate, of course, requires more detailed consideration.

In Fig. 2a, the area of sterile neutrino parameters, determined by the exclusions from experiments Troitsk, KATRIN, BEST and DANSS is highlighted in blue. The result of the Neutrino-4 experiment is inside this area.

The comparison with the results of other reactor neutrino experiments DANSS [13], NEOS [14], PROSPECT [15] and STEREO [16] can be found in work [8]. The experiments with the nuclear power plant as a neutrino source have significantly less sensitivity to the big values of the parameter $\Delta m_{14}^2$ due to the large size of the reactor core. The illustration of the comparison with the reactor experiments is presented in Fig.2b [8]. The sensitivity of the experiments STEREO [16] and PROSPECT [15] is yet not sufficient to exclude the result of the Neutrino-4 experiment, which has twice better sensitivity (Fig.2b). It is especially important, that in the Neutrino-4 experiment we observe the oscillation process directly in the measurements. Other experiments results in the exclusion with less experimental sensitivity. Further improvement of all listed experiments is possible and at some point, the answer for the problem of existence of the sterile neutrino would be revealed. Figure 3 demonstrates comparison of the new BEST experiment result with the result of Neutrino-4 experiment.

The result of the KATRIN [17,18] experiment does not exclude the Neutrino-4 area (Fig.2a). The GERDA [19] experiment requires special attention, since it is aimed at search for Majorana type neutrino mass. Currently, the limit on the Majorana mass obtained in the GERDA experiment for normal mass hierarchy is one standard deviation less than the prediction of the Majorana mass derived from the result of the Neutrino-4 experiment (dark grey area on Fig.2a). If in the future the limit of the Majorana mass of the experiment on double beta decay is lowered, and the result of the Neutrino-4 experiment is confirmed, this will close the hypothesis that the neutrino is the Majorana type particle.

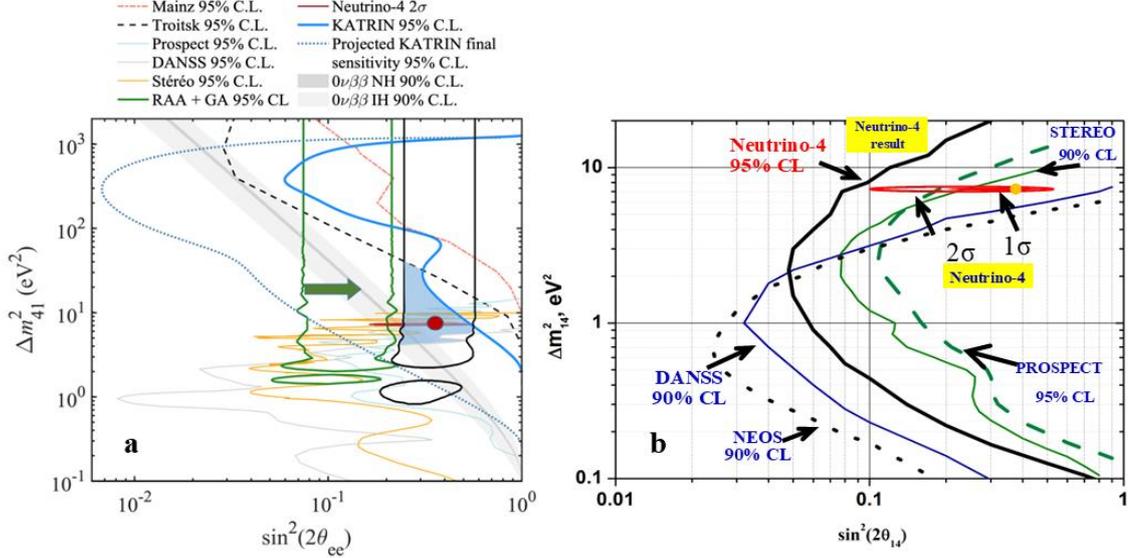

Fig.2. a – Comparison of results of the Neutrino-4 experiment with results of other experiments [16], the BEST result combined with the previous SAGE and GALLEX results area is added, b – Comparison of sensitivities of the experiments: Neutrino-4, STEREO, PROSPECT, DANSS and NEOS [8].

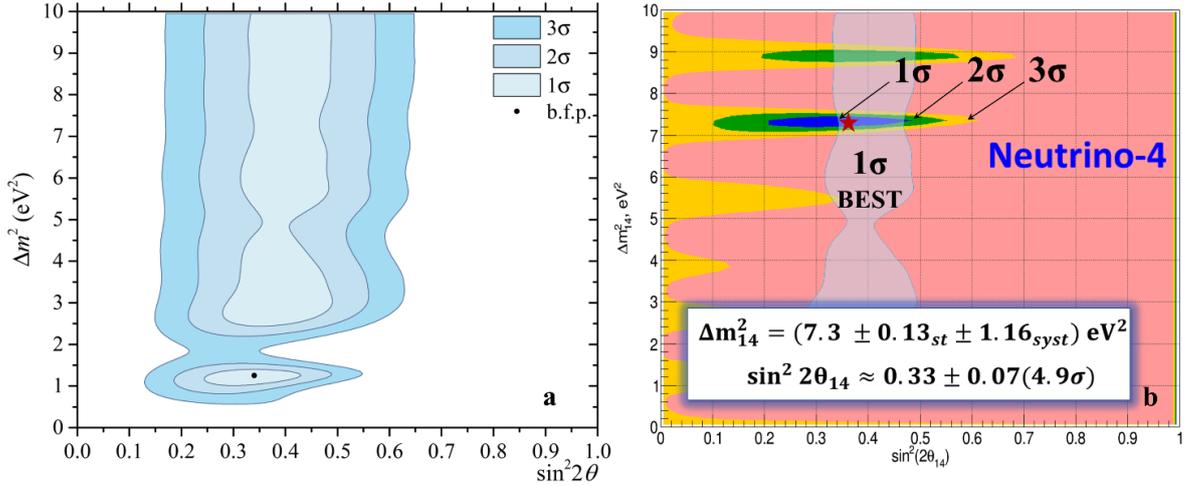

Fig. 3. Comparison of the BEST experiment result and the result of Neutrino-4 experiment. a – confidence levels of the BEST result combined with the previous SAGE and GALLEX results with best fit point, b –result of overlapping 1σ area of the BEST experiment and $\Delta m^2_{14}, \sin^2 2\theta_{14}$ plane with CL of Neutrino-4 and combined analysis parameters of the 2 experiments.

**4. The relations of the different oscillation process probabilities in the 3+1 neutrino model**

The discussion of the neutrino experiments at accelerators in context of the Neutrino-4 result requires the brief description of the 3+1 neutrino model.

$$\begin{bmatrix}\nu_e\\\nu_\mu\\\nu_\tau\\\nu_s\end{bmatrix} = \begin{bmatrix}U_{e1} & U_{e2} & U_{e3} & U_{e4}\\U_{\mu 1} & U_{\mu 2} & U_{\mu 3} & U_{\mu 4}\\U_{\tau 1} & U_{\tau 2} & U_{\tau 3} & U_{\tau 4}\\U_{s1} & U_{s2} & U_{s3} & U_{s4}\end{bmatrix}\begin{bmatrix}\nu_1\\\nu_2\\\nu_3\\\nu_4\end{bmatrix} \quad (4)$$

$$|U_{e4}|^2 = \sin^2(\theta_{14})$$
$$|U_{\mu 4}|^2 = \sin^2(\theta_{24})\cdot\cos^2(\theta_{14}) \quad (5)$$
$$|U_{\tau 4}|^2 = \sin^2(\theta_{34})\cdot\cos^2(\theta_{24})\cdot\cos^2(\theta_{14})$$

$$P_{\nu_e\nu_e} = 1 - 4|U_{e4}|^2(1-|U_{e4}|^2)\sin^2\left(\frac{\Delta m^2_{14}L}{4E_{\nu_e}}\right) = 1 - \sin^2 2\theta_{ee}\sin^2\left(\frac{\Delta m^2_{14}L}{4E_{\nu_e}}\right) \quad (6)$$

$$P_{\nu_\mu\nu_\mu} = 1 - 4|U_{\mu 4}|^2\left(1-|U_{\mu 4}|^2\right)\sin^2\left(\frac{\Delta m^2_{14}L}{4E_{\nu_\mu}}\right) = 1 - \sin^2 2\theta_{\mu\mu}\sin^2\left(\frac{\Delta m^2_{14}L}{4E_{\nu_\mu}}\right) \quad (7)$$

$$P_{\nu_\mu\nu_e} = 4|U_{e4}|^2|U_{\mu 4}|^2\sin^2\left(\frac{\Delta m^2_{14}L}{4E_{\nu_e}}\right) = \sin^2 2\theta_{\mu e}\sin^2\left(\frac{\Delta m^2_{14}L}{4E_{\nu_e}}\right) \quad (8)$$

One can see that if the mixing angles are small then the new matrix elements $U_{e4}, U_{\mu4}, U_{\tau4}$ can be determined by measuring the mixing amplitudes. The probabilities of the various oscillation processes are listed in expressions 6 to 8. Eq.6 is the probability of disappearance of the electron neutrino due to oscillation in the sterile state. Eq.7 is the probability of disappearance of the muon neutrino due to oscillation in the sterile state. Eq.8 is the probability of the oscillation of the muon neutrino into the electron neutrino through the sterile state. The oscillation amplitudes in these processes are:

$$\sin^2 2\theta_{ee} \equiv \sin^2 2\theta_{14} \quad (9),$$

$$\sin^2 2\theta_{\mu\mu} = 4\sin^2\theta_{24}\cos^2\theta_{14}(1 - \sin^2\theta_{24}\cos^2\theta_{14}) \approx \sin^2 2\theta_{24} \quad (10),$$

$$\sin^2 2\theta_{\mu e} = 4\sin^2\theta_{14}\sin^2\theta_{24}\cos^2\theta_{14} \approx \tfrac{1}{4}\sin^2 2\theta_{14}\sin^2 2\theta_{24} \quad (11).$$

The relatively small values of the mixing angles open the way to simplify the expressions without significant loss of accuracy. The new parameters in the simplified expressions (squared sinus of the mixing angles) can be directly measured in the experiments. The listed above relations of the parameters allow us to compare results of the different types of oscillation experiments.

In the 3+1 neutrino model with one sterile neutrino the oscillation length should be the same for all processes and it determined by the parameter $\Delta m_{14}^2$. Also, the amplitudes of the electron and muon neutrino oscillations in the disappearance processes determine the amplitude of the of the appearance of the electron neutrinos in the muon neutrino flux $\sin^2 2\theta_{\mu e} \approx \tfrac{1}{4}\sin^2 2\theta_{14}\sin^2 2\theta_{24}$. This expression is very important for the problem of validation of the 3+1 neutrino model. It has rather simple interpretation. The appearance of the electron neutrinos in the muon neutrino flux is the second order process, which can be considered as oscillation of the muon neutrino into the sterile state and following oscillation of the sterile neutrino into the electron neutrino.

The experiments that observe the effects, which indicates the oscillation into the sterile state in the flux of electron neutrinos are Neutrino-4, RAA and GA. The effects, which can be interpreted as the oscillation into the sterile state in the muon neutrino flux yet have been observed with relatively low CL in the IceCube experiment. And finally, the experiments which claimed to observe the appearance of the electron neutrinos in the muon neutrino flux are MiniBooNE and LSND.

## 5. Comparison of the Neutrino-4 results with the results of the IceCube, MiniBooNE and LSND experiments

Figure 4 illustrates the comparison of the Neutrino-4 experimental result with the values obtained in the IceCube, MiniBooNE and LSND experiments in context of the 3+1 neutrino model. The best fit value obtained in the IceCube experiment [20] is $\Delta m_{14}^2 = 4.47^{+3.53}_{-2.08} eV^2$; $\sin^2(2\theta_{24}) = 0.10^{+0.10}_{-0.07}$. The values of the parameter $\Delta m_{14}^2$ which determines the oscillation length obtained in the IceCube and Neutrino-4 are in agreement within one standard deviation, but we should mention, that the accuracy of the IceCube value is rather small and insufficient for reliable comparison. The figure 4 depicts the direct verification of the equation $\sin^2 2\theta_{\mu e} \approx \tfrac{1}{4}\sin^2 2\theta_{14}\sin^2 2\theta_{24}$ where we use values $\sin^2 2\theta_{14} \approx 0.33 \pm 0.07$ (4.9σ) (Neutrino-4 and GA), and $\sin^2 2\theta_{24} = 0.10^{+0.10}_{-0.07}$ obtained in the IceCube experiment. The resulting value $\sin^2 2\theta_{\mu e} \approx \tfrac{1}{4}\sin^2 2\theta_{14}\sin^2 2\theta_{24} = 0.009^{+0.011}_{-0.008} \approx 0.001 \div 0.020$ is marked in fig.4 by the 1σ contour. We can conclude that at current precision level the values of mixing angles obtained in the MiniBooNE, LSND, Neutrino-4 and IceCube are in in good agreement within the 3+1 framework. Yet, the increase of accuracy of the neutrino experiments is required.

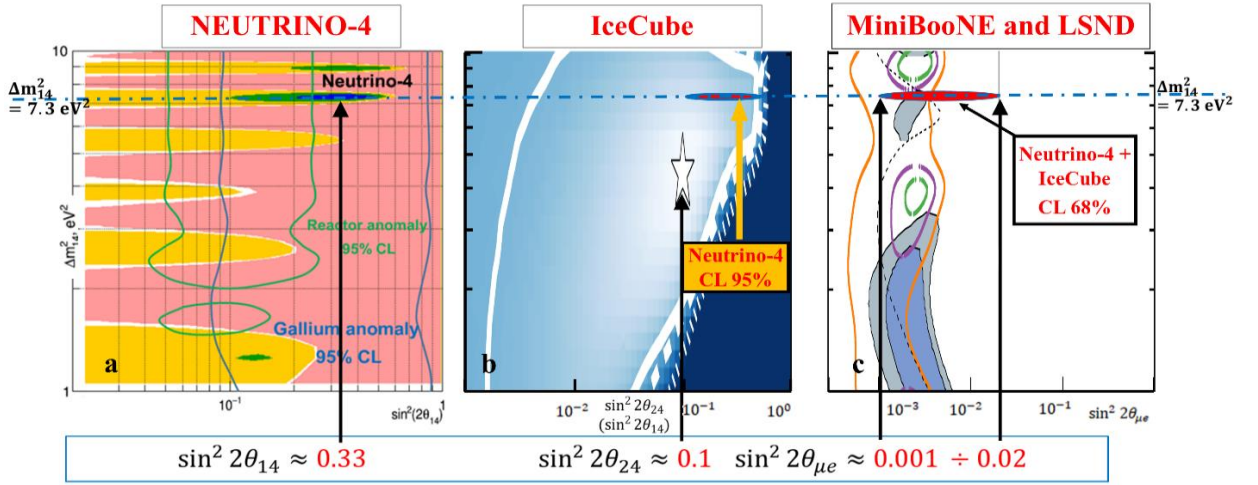

Fig.4. a – The comparison of Neutrino-4 data with the IceCube results, b – Neutrino-4 data in comparison with the MiniBooNE and LSND accelerator experiments on the $(\sin^2 2\theta_{\mu e}, \Delta m_{14}^2)$ plane and the test of the relation $\sin^2 2\theta_{\mu e} \approx \tfrac{1}{4}\sin^2 2\theta_{14}\sin^2 2\theta_{24}$

## 6. The prediction of the effective mass of the electron neutrino derived from the Neutrino-4 results and comparison with the results of the experiments on measurement of the neutrino mass: KATRIN and GERDA

The oscillation parameters obtained in the Neutrino-4 experiment can be applied to estimate the effective electron neutrino mass in the 3+1 neutrino model, using the well-known ratio [21,22]:

$$m_{\nu_e}^{\text{eff}} = \sqrt{\sum m_i^2 |U_{ei}|^2}\,; \quad \sin^2 2\theta_{14} = 4|U_{14}|^2;$$

The sum of neutrino masses $\sum m_\nu = m_1 + m_2 + m_3$ is limited by the cosmological researches by the value $0.54 \div 0.11$ eV [21].Therefore, if we consider $\Delta m_{14}^2 = 7.3$ eV$^2$, then we can simplify the 3+1 framework masses to $m_4^2 \approx 7.3$ eV$^2$, and $m_1^2, m_2^2, m_3^2 \ll m_4^2$. The effective mass of the electron neutrino can be calculated using the equation:

$$m_{\nu_e}^{\text{eff}} \approx \sqrt{m_4^2 |U_{e4}|^2} \approx \tfrac{1}{2}\sqrt{m_4^2 \sin^2 2\theta_{14}}.$$

The mass of the sterile neutrino can be estimated as: $m_4 = (2.70 \pm 0.22)$eV. The parameter $\sin^2 2\theta_{14} \approx 0.33 \pm 0.07(4.9\sigma)$ (Neutrino − 4 and GA) and, more importantly, the obtained in the Neutrino-4 experiment value $\Delta m_{14}^2 \approx (7.3 \pm 1.17)$eV$^2$, can be applied to estimate the mass of the electron neutrino: $m_{4\nu_e} = (0.82 \pm 0.16)$eV or in terms of squared mass $m_{4\nu_e}^2 = (0.67 \pm 0.26)$ eV$^2$

The indication if the neutrino model used to calculate the effective mass is essential to the interpretation of the result. The KATRIN experiment is aimed at the direct measurement of the neutrino mass using the process T → He$_3$ + $e^-$ + $\bar\nu$, near the threshold 18.6keV. The effective electron neutrino mass obtained by the KATRIN collaboration is $m_{3\nu_e}^2 = 0.26 \pm 0.34$ and the upper limit is $m_{3\nu_e}^{\text{eff}} \leq 0.8$ eVC.L.90% [18], but it is calculated in the three neutrino framework with 3x3 unitary PMNS matrix and in assumption that the mass differences $\Delta m_{12}^2$ and $\Delta m_{13}^2$ are significantly less than 1 eV$^2$, therefore it cannot be directly compared to the result obtained in the analysis of the Neutrino-4 data.

The analysis of the KATRIN data within the 3+1 framework can be found in work [17]. Current experimental result can be used only to construct the exclusion contour at the ($\Delta m_{14}^2$; $\sin^2 2\theta_{14}$) plane, which is shown in Fig.2a. The excluded region does not close the Neutrino-4 result. The estimations within the 3+1 model in assumption $m_1^2, m_2^2, m_3^2 \ll m_4^2$ are listed in the Table 1 in the Neutrino-4 column, while the KATRIN column list the results of data analysis with hypothesis of only three neutrinos without sterile state.

In principle, the KATRIN experimental data can be used in analysis with three free parameters: mixing angle $U_{e4}$, mass $m_4^2$ and the effective mass of the 3 light neutrinos $m_{3\nu}^2 = \frac{\sum_{i=1}^{3} m_i^2}{1-U_{e4}^2}$. Currently, the amount of collected data is insufficient to carry out such analysis without additional assumptions. But the parameters of the fourth neutrino obtained in the Neutrino-4 experiment $\sin^2 2\theta_{14} \approx 0.33 \pm 0.07(4.9\sigma)$ and $m_4^2 \approx 7.3$ eV$^2$ can be used as such additional parameters. So, we hope our colleagues from the KATRIN collaboration would present such analysis in one of their future publications.

Using the value of $\sin^2 2\theta_{24}$ obtained in the IceCube experiment we can estimate the mass of the muon neutrino: $m_{\nu_\mu}^{\text{eff}} = (0.41 \pm 0.25)$eV. Finally, we can use the limit $\sin^2 2\theta_{34} \leq 0.21$ obtained in the IceCube experiment to calculate the upper limit of the tau neutrino mass: $m_{\nu_\tau}^{\text{eff}} \leq 0.60$eV.

The effective Majorana mass of the electron neutrino which is the parameter in the analysis of the double β-decay experiments is determined by the ratio:

$$m(0\nu\beta\beta) = \left|\sum_i U_{ei}^2 m_i\right|$$

Which in the framework 3+1 and $m_1, m_2, m_3 \ll m_4$ can be simplified to $m(0\nu\beta\beta) \approx m_4 U_{14}^2$. The quantitative result of this ratio is $m(0\nu\beta\beta) = (0.25 \pm 0.07)$eV. The strictest limit on the Majorana mass is the result of the GERDA experiment [19]. This experiment is aimed at the measuring of the half-life of the isotope, which is the function of the Majorana mass in the model with neutrinoless double beta decay. The upper limit of the half-life yields the upper limit of the Majorana mass:

- Lower 90% CL limit: $T_{1/2}^{0\nu} > 1.8 \times 10^{26}$ yr
- Converted into upper limit on: $m_{\beta\beta} < [79 - 180]$ meV

The value obtained with the Neutrino-4 oscillation parameters is $m(0\nu\beta\beta) = (0.25 \pm 0.07)$eV and it is three times bigger than the lowermost limit claimed by the GERDA experiment. It is a significant discrepancy, but it is too early to make reliable conclusions. Therefore, further increase of accuracy of the experiments in search for the double neutrinoless beta decay is important for understanding of the neutrino physics.

**Table 1**
The estimation of the electron neutrino effective mass

|  | Neutrino-4 | KATRIN | GERDA $m(0\nu\beta\beta)$ |
|---|---|---|---|
| Effective mass and mass squared ($m_{\nu_e}^{\text{eff}}, (m_{\nu_e}^{\text{eff}})^2$) | $m_{4\nu_e}^{\text{eff}} = (0.82 \pm 0.16)$eV  $(m_{4\nu_e}^{\text{eff}})^2 = (0.67 \pm 0.26)$eV$^2$ | $m_{3\nu_e}^{\text{eff}} < 0.8$ eV (90%)  $(m_{3\nu_e}^{\text{eff}})^2 = (0.26 \pm 0.34)$eV$^2$ |  |
| Maiorana mass ($m(0\nu\beta\beta)$) | $m(0\nu\beta\beta) = (0.25 \pm 0.07)$eV |  | $m_{\beta\beta} < [0.079 - 0.180]$eV |

### 7. PMNS matrix within the 3+1 neutrino model

The PMNS matrix of the 3+1 model with sterile neutrino parameters measured in the Neutrino -4 experiment, the RAA and GA and in the IceCube experiment is shown below:

$$U_{PMNS}^{(3+1)} = \begin{pmatrix} 0.824^{+0.007}_{-0.008} & 0.547^{+0.011}_{-0.011} & 0.147^{+0.003}_{-0.003} & 0.302^{+0.034}_{-0.034} \\ 0.409^{+0.036}_{-0.060} & 0.634^{+0.022}_{-0.065} & 0.657^{+0.044}_{-0.014} & 0.152^{+0.080}_{-0.057} \\ 0.392^{+0.025}_{-0.048} & 0.547^{+0.056}_{-0.028} & 0.740^{+0.012}_{-0.048} & < 0.222 \\ < 0.24 & < 0.30 & < 0.26 & > 0.91 \end{pmatrix}$$

The limits on the parameters $U_{si}$ are calculated using the unitarity as an assumption that sum of squared elements in the column do not exceed one for more than one standard deviation. Below we present the mixing scheme of the active neutrinos and a sterile neutrino in the flavor basis for direct and inverse mass hierarchy (fig.5). Though we should notice, that in both cases the main effect is due to the difference in masses between the sterile and SM neutrinos.

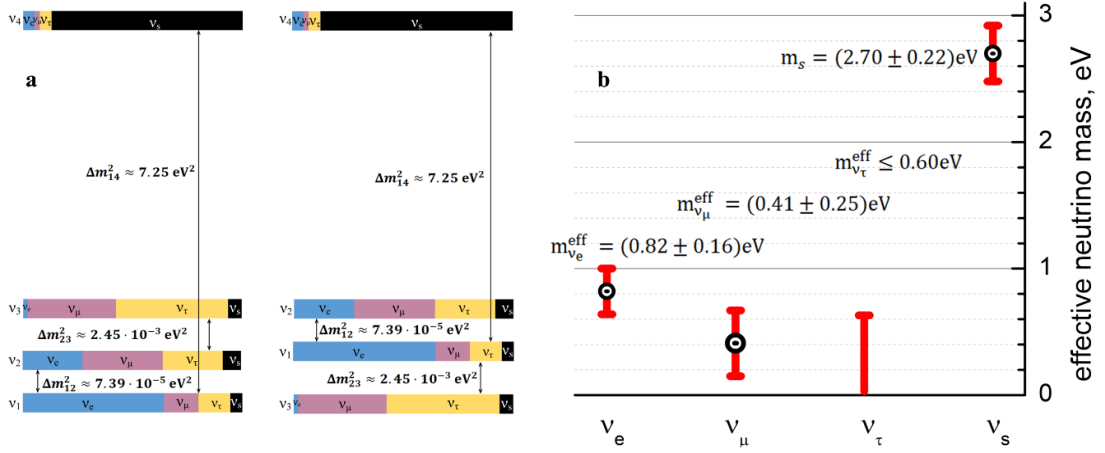

Fig.5. a – The mixing scheme of flavor neutrinos in the 3+1 neutrino model with direct and inverse mass hierarchy, b – The estimations of the effective neutrino masses.

## 8. The cosmological restrictions and the Neutrinoo-4 results.

Finally, one has to take into account the restriction on the number of the active neutrino flavors and sum of active neutrino masses from the cosmology, that can affect the introduction of the fourth neutrino. In various models the massive neutrinos are considered to take significant part in the early Universe processes, baryon asymmetry of the Universe and the Dark Matter problem. However, the light sterile neutrino with a mass of the several eV and small mixing with active neutrinos can exist without any significant impact on the structure of the Universe. Such neutrinos do not thermalize in the primordial plasma due to extremely weak interaction with matter and leave it at early stage. Yet in that process, the sterile neutrino decreases the mass of the primordial plasma and hence accelerate the expansion of the Universe.

## 9. Conclusion

The 3+1 neutrino model is considered in the context of the new result of the Neutrino-4 experiment, the direct observation of the oscillation effect at 2.9$\sigma$ CL at parameter region $\Delta m_{14}^2 = (7.3 \pm 0.13_{st} \pm 1.16_{syst})$ eV$^2$ and $\sin^2 2\theta_{14} = 0.36 \pm 0.12_{stat}(2.9\sigma)$, using the LSND anomaly, MiniBooNE anomaly, reactor anomaly, reactor antineutrino anomaly (RAA), and gallium anomaly observed in experiments with radioactive sources GALLEX/GNO, SAGE and BEST.

Meanwhile, it should also be noted recently published in arXiv result of the BEST experiment confirms value of the mixing angle obtained in Neutrino-4 experiment.

We suggest two criteria to experimentally verify the 3+1 neutrino model with one sterile neutrino. The first criterion - the oscillation frequency of all oscillation processes at short distance is the same, i.e. it is determined by the parameter $\Delta m_{14}^2$. The measured values of $\Delta m_{14}^2$ from the Neutrino-4 and IceCube experiments are in agreement within one standard deviation, however the accuracy of the IceCube experiment should be increased. The second criterion – the validity of the ratio $\sin^2 2\theta_{\mu\nu} \approx \frac{1}{4}\sin^2 2\theta_{14} \sin^2 2\theta_{24}$ for the mixing amplitudes of the oscillation processes. At the current precision level, the mixing angles of the experiments MiniBooNE, LSND, Neutrino-4 and IceCube are in agreements within the 3+1 framework. Yet, it is essential to increase the accuracy of the experiments to obtain the sufficient CL in the combined analysis.

The oscillation parameters obtained in the Neutrino-4 experiment can be applied to calculate the estimation of the effective electron neutrino mass within 3+1 neutrino model: $m_{4\nu_e}^{eff} = (0.82 \pm 0.16)$eV. The calculated mass is at the same level as the limit obtained in the KATRIN experiment [18] $m_{3\nu_e}^{eff} \leq 0.8$ eVC.L.90%.

In the considered framework we can estimate the majorana neutrino mass: $m(0\nu\beta\beta) = (0.25 \pm 0.07)$eV. It should be compared with the results of the GERDA experiment: $m_{\beta\beta} < [0.079 - 0.180]$eV. The estimation based on the parameters of the sterile neutrino is two times bigger than the current upper limit obtained in the GERDA experiment. Therefore, in order to make reliable conclusions the accuracy of the experiments should be increased several times. The further decrease of the majorana mass limit and the confirmation of the Neutrino-4 result could in future close the hypothesis of the majorana neutrino.

Finally, the available experimental data of the considered experiments we can build the PMNS matrix of the 3+1 model. The value $U_{e4} = 0.302^{+0.034}_{-0.034}$ is measured with high precision. The value $U_{\mu4} = 0.152^{+0.080}_{-0.057}$ is measured only up to 2 $\sigma$ level. Other new values of the matrix have only upper limit, estimated using the unitarity of the matrix. Using the calculated matrix one can get the scheme of the neutrino mixing in the 3+1 neutrino model. That scheme reveals that the mass of the sterile neutrino determines the masses of 3 active flavor neutrinos. The obtained neutrino effective mass hierarchy is: $m_{\nu_e}^{eff} = (0.82 \pm 0.16)$eV, $m_{\nu_\mu}^{eff} = (0.41 \pm 0.25)$eV, $m_{\nu_\tau}^{eff} \leq 0.60$eV.

In summary, the performed analysis provides quite interesting generalizations and indications of the validity of the 3 + 1 neutrino model with one sterile neutrino.

### Acknowledgements

The work was supported by the Russian Science Foundation under Contract No. 20-12-00079.